# TMA: Tera-MACs/W Neural Hardware Inference Accelerator with a Multiplier-less Massive Parallel Processor

Hyunbin Park, Dohyun Kim, and Shiho Kim

*Abstract*—Computationally intensive Inference tasks of Deep neural networks have enforced revolution of new accelerator architecture to reduce power consumption as well as latency. The key figure of merit in hardware inference accelerators is the number of multiply-and-accumulation operations per watt (MACs/W), where, the state-of-the-arts MACs/W remains several hundred's Giga-MACs/W. We propose a Tera-MACS/W neural hardware inference Accelerator (TMA) with 8-bit activations and scalable integer weights less than 1-byte. The architecture's main feature is configurable neural processing element for matrix-vector operations. The proposed neural processing element has Multiplier-less Massive Parallel Processor to work without any multiplications, which makes it attractive for energy efficient high-performance neural network applications. We benchmark our system's latency, power, and performance using Alexnet trained on ImageNet. Finally, we compared our accelerator's throughput and power consumption to the prior works. The proposed accelerator significantly outperforms the state-of-the-art in terms of energy and area efficiency achieving 2.3 TMACS/W@1.0 V, 65 nm CMOS technology.

*Index Terms*—Enter key words or phrases in alphabetical order, separated by commas. For a list of suggested keywords, send a blank e-mail to keywords@ieee.org or visit http://www.ieee.org/organizations/pubs/ani_prod/keywrd98.txt

## I. INTRODUCTION

DEEP Neural Networks (DNNs) have driven artificial intelligence to outperform human's recognition of images such as animal, object, and terrain [1]. The strong demand on parallel processing hardware to process such multiply-and-accumulation (MAC)-intensive computation in the DNNs has revolutionized the computer industry. In particular, attempts have been made to incorporate such parallel processors into embedded systems. However, embedded systems for ultralow power applications such as autonomous vehicles, drones, and smart robots are still facing hard power constraint. To meet power budget of such embedded systems, several studies have employed a strategy that runs inference on hardware accelerator in embedded systems, while running training on General-Purpose computing on Graphics Processing Units (GP-GPUs) [2]-[4].

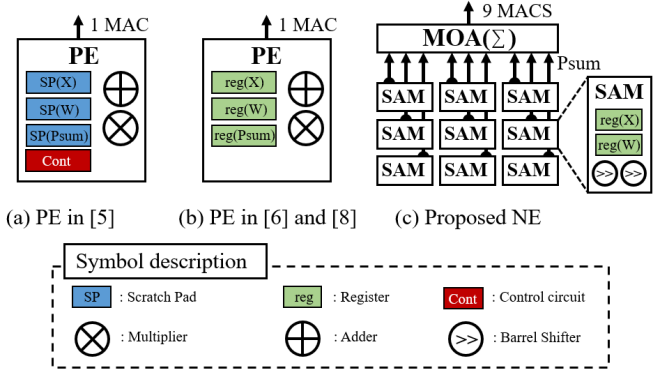

Fig. 1. Block diagram of a processing element of the previous works (a) Eyeriss [5], (b) ConvNet [6], DSIP [8], and (c) the proposed work.

The key figure of merit in hardware accelerators is the number of multiply-and-accumulation operations per watt (MACs/W); it is important not only to increase throughput but also to decrease power consumption. However, MACs/W of the state-of-the-arts remains several hundred's Giga-MACs/W [5]-[8]. To expand application area of the neural processing, the performance should be improved by at least one order of magnitude.

The throughput of the hardware accelerator depends on how many MAC operations are performed simultaneously in parallel. The previous works [5], [6], and [8] comprise 168, 256, and 64 processing elements (PEs), respectively. Each PE in [5] contains not only energy-intensive 16-bit multiplier and 16-bit binary adder but also three scratchpad memories, as shown in Fig. 1 (a). Therefore, it limits the number of PEs can be mounted in an accelerator. [6] and [8] does not employ scratchpad memories in each PE, which reduces the circuit complexity of a PE compared to [5]. However, [6] and [8] assign an input or a weight into multiple PEs, which makes computation in Fully-Connected (FC) layers disadvantageous; since, in the computation in FC layers, the elements of the

"This research was supported by the MSIT(Ministry of Science and ICT), Korea, under the "ICT Consilience Creative Program" (IITP-2019-2017-0-01015) supervisedby the IITP(Institute for Information & communications Technology Planning&Evaluation) "

The authors are with the School of Integrated Technology, Yonsei University, Seoul, Korea (e-mail: bin9000@yonsei.ac.kr; kimdh5032@yonsei.ac.kr; shiho@yonsei.ac.kr).

flattened weights and the elements of the flattened inputs are multiplied with a one-to-one correspondence way.

To reduce circuit complexity of PEs, [9]-[11] reduced bit-width of inputs and weights lower than 16-bit. [11] refers that the quantization of bit-width of both weights and activations into 8-bit integers in inference causes additional Top-1 errors in Resnet [1], VGG19 [12], and AlexNet [13] less than 0.2 % compared to Floating Point 32-bit (FP32).

To further reduce circuit complexity of PEs, hardware accelerators [14]-[15] adopt bit-shift-based-multipliers instead of conventional multipliers for MAC operations. In SiMul [14], a bit-shifter produces one partial product per cycle and then it is accumulated. Bit fusion [15] decomposes a partial product with the trapezoidal format into multiple regions of the smaller trapezoidal portion. The decomposed partial products are aggregated by binary adders, and a bit-shifter apply arithmetic shift operation to the summed output. Both studies additionally allow bit-width of weight to be flexibly adjusted.

We propose a Tera-MACS/W neural hardware inference accelerator (TMA) with a multiplier-less massive parallel processor with 8-bit integer activations and integer weights less than 1-byte. The main objective of this paper is to achieve Tera-MACS/W. To achieve that, we propose a Neural Element (NE), a basic unit of a computational engine, that computes 9 MAC operations in parallel with 9 multiplier-less and scratchpad-less shift-and-multiplication (SAM) circuits and a multi-operand adder (MOA) circuit, as shown in Fig. 1 (c). Additionally, to further reduce power consumption of the NE, the proposed TMA accelerator permits multiplication errors in some cases. This paper also proposes a configurable architecture with scalable filter size that performs computations in both convolutional (Conv) and FC layers in a massive parallel scheme.

## II. ARCHITECTURE OF THE TMA ACCELERATOR

The proposed NE of the TMA accelerator employs two-barrel shifters to multiply an integer input and an integer weight. In the first sub-section, we propose a quantization technique to compute an arithmetic multiplication using two-barrel shifters. The second sub-section describes an architecture of the proposed TMA accelerator and operational principle for computing convolution. Lastly, in the third sub-section, we explain a configuration scheme of the proposed TMA accelerator that supports variable filter sizes and computation in FC layers

### A. Quantization with 2N Partial Sub-integers

The proposed TMA accelerator partitions a product of an integer input X and an integer weight w into 2N signed $2^n$, as indicated in Eq. (1), where the partitioned integers are called partial sub-integers (PSIs).

$$w \cdot X = \sum_{k=1}^{N}(s1_k \cdot 2^{n1_k} \cdot X + s2_k \cdot 2^{n2_k} \cdot X) \quad (s1_k, s2_k \in -1,0,1)$$
$$= \sum_{k=1}^{N}(PSI1_k + PSI2_k) \quad (1)$$

As illustrated in Fig. 1 (c), two-barrel shifters in the SAM block produce two partial sub-integers PSI1 and PSI2, and they are accumulated based on the bit-width of weights. We partition the product of X and a 5-bit integer weight into two PSIs (i.e., N is 1 in Eq. (1)), and the product of X and 8-bit integer weight into four PSIs (i.e., N is 2 in Eq. (1)). However, partitioning of the 5-bit weight into two PSIs causes multiplication errors of approximately 9 % only when the weight is $-13$, $-11$, 11, or 13, while partitioning into four PSIs does not cause an error. This is summarized in Table I.

TABLE I
MULTIPLICATION ERROR IN TERMS OF NUMBER OF PARTIAL SUB-INTEGERS

| Number of partitions | Precision of weight | Multiplication error in worst case | Inference accuracy degradation (Top-1) compared to FP32 | |
|---|---|---|---|---|
| | | | LeNet-5 (MNIST) | AlexNet (ImageNet) |
| 2 PSIs | INT5 | ~ 9 % | 0 % | 3.9 % |
| 4 PSIs | INT8 | 0 % | 0 % | 0.1 % |

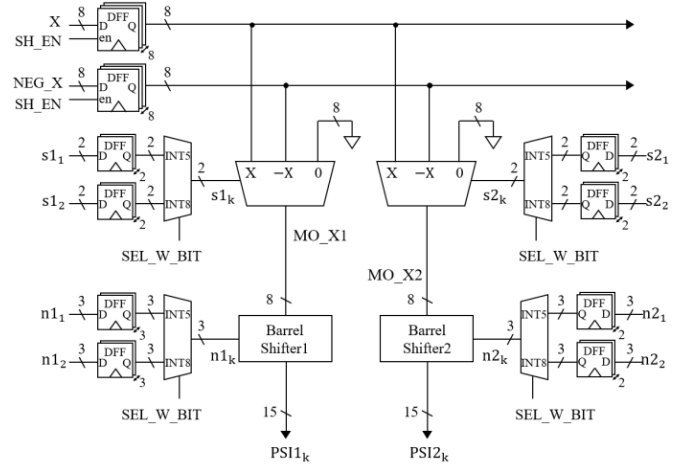

Fig. 2. Block diagram of the SAM block of the proposed TMA accelerator which is illustrated in Fig. 1 (c).

To show how the proposed quantization degrade the inference accuracy, we trained LeNet-5 [16] and AlexNet [13] with the proposed quantization with training MNIST [16] and ImageNet [17] databases, respectively. The proposed quantization of weight into both INT5 and INT8 does not degrade the inference accuracy in LeNet-5. However, the proposed quantization into INT5 reduces the inference accuracy by 3.9 % in Top-1 error. Therefore, we suggest that users of the TMA accelerator select the bit-width of the weight based on neural networks.

### B. Architecture of the Proposed TMA Accelerator

This subsection describes an architecture of the proposed NE, a basic multiplier-less computational engine. We will also describe the NE array and overall system architecture in this subsection.

The NE comprises the MOA block and 9 SAM blocks as shown in brief illustration in Fig. 1 (c). Fig 2 shows the block diagram of the SAM block. As indicated in in Eq. (1), the SAM receives decomposed elements of a weight, which are $s1_1$, $s1_2$, $s2_1$, $s2_2$, $n1_1$, $n1_2$, $n2_1$, and $n2_2$. It also receives an 8-bit positive input X, an 8-bit negatized input NEG_X, and 8-bit zero, among which 3-1 mux circuits select one based on $s1_k$ and SEL_W_BIT signals. The output signals of the mux, MO_Xs, are shifted based on $n1_k$ or $n2_k$ bits to produce two PSIs.

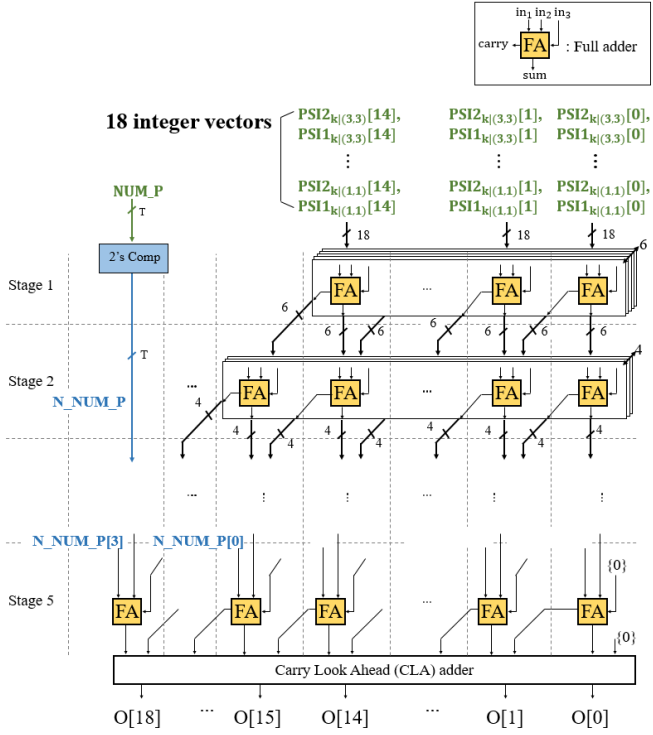

Fig. 3. Block diagram of the proposed MOA18. It aggregates 18 signed integers.

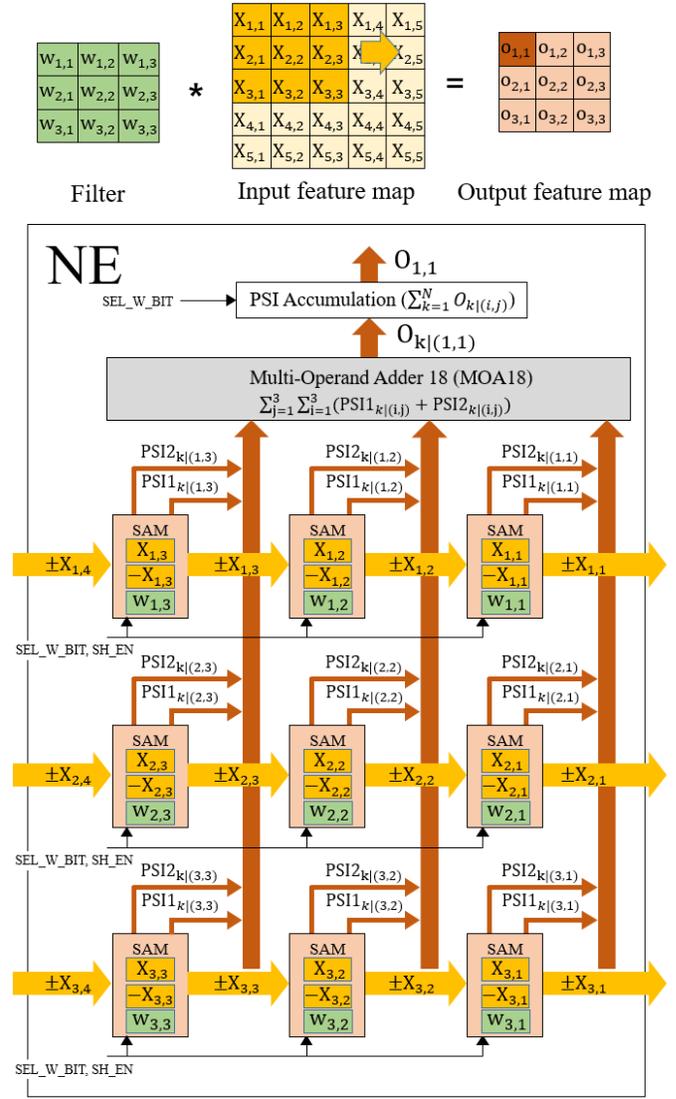

The MOA18 block collects 18 PSIs from 9 SAMs in the NE, and aggregates them with massive-parallel scheme. Fig. 3 shows block diagram of the proposed MOA18 block, where the number of inputs is 18. The operational principle of the MOA block is similar with the Wallace tree adder [18]. The MOA block groups three bits in the same column of vectors along the vertical direction into one group. After that, full adder arrays in each stage reduce the number of input vectors by two-thirds, and simultaneously expands bit-width of them by one bit. At the output of the stage 5, the number of vectors is reduced to two. Finally, the two vectors are added by a Carry Look Ahead (CLA) adder.

To permit calculation of the sum of numbers with individual signs, a sign-extension of inputs of the MOA to 18-bits would be required, where the 18-bit is the bit-width of the output of the MOA. However, the sign-extension increases circuit area by approximately 21 %. Instead of sign-extending, we negatize NUM_P, where NUM_P is the number of negative partial sub-integers. Then, we add it to from 15-bit to 18-bit in Stage 5. It requires only 2's complement circuit, and allows the calculation of the sum of numbers with individual signs. The principle of the simplified sign-extension is described in Appendix with Fig. A1. The proposed MOA18 circuit reduces total gates by 36 % compared to the 18 hierarchical CLA adders [19].

Figure 4 shows block diagram of the NE and its computation of 3×3 convolution. The stored positive and negative inputs X

Fig. 4. Block diagram of the proposed NE and its computation of 3×3 convolution.

and –X in SAMs are connected in the horizontal direction. The inputs slides to the horizontal direction based on the SH_EN signal as the filter sweeps. Weights are stationarily held in registers with decomposed format in SAMs as the filter sweeps.

The output of MOA18 is the convolutional result of 3×3 patch with 2 PSIs. As indicated in Eq. (1), NE should support accumulation of PSIs to flexibly adjust bit-width of weights. The PSI Accumulation block accumulates the PSIs based on the SEL_W_BIT signal by accumulating the output of MOA18.

The proposed TMA accelerator adopts 4 × 4 × 16 NE array for neural processing. With this NE array, 2,304 MACs can be computed in parallel in maximum. In case of computing four 3 × 3 × D convolutions as illustrated in Fig. 5, weights of the four 3 × 3 × 64 filters are assigned to NEs in the same column. The four filters share a single input feature map. Therefore, inputs are slide through NE array in the horizontal direction. The 64 outputs of 64 NEs in the same column are collected by MOA66 block, which aggregates the Partial sum of computation in Conv or FC layers (Psum) loaded from SRAM, Bias loaded from SRAM, and the 64 outputs produced from 64

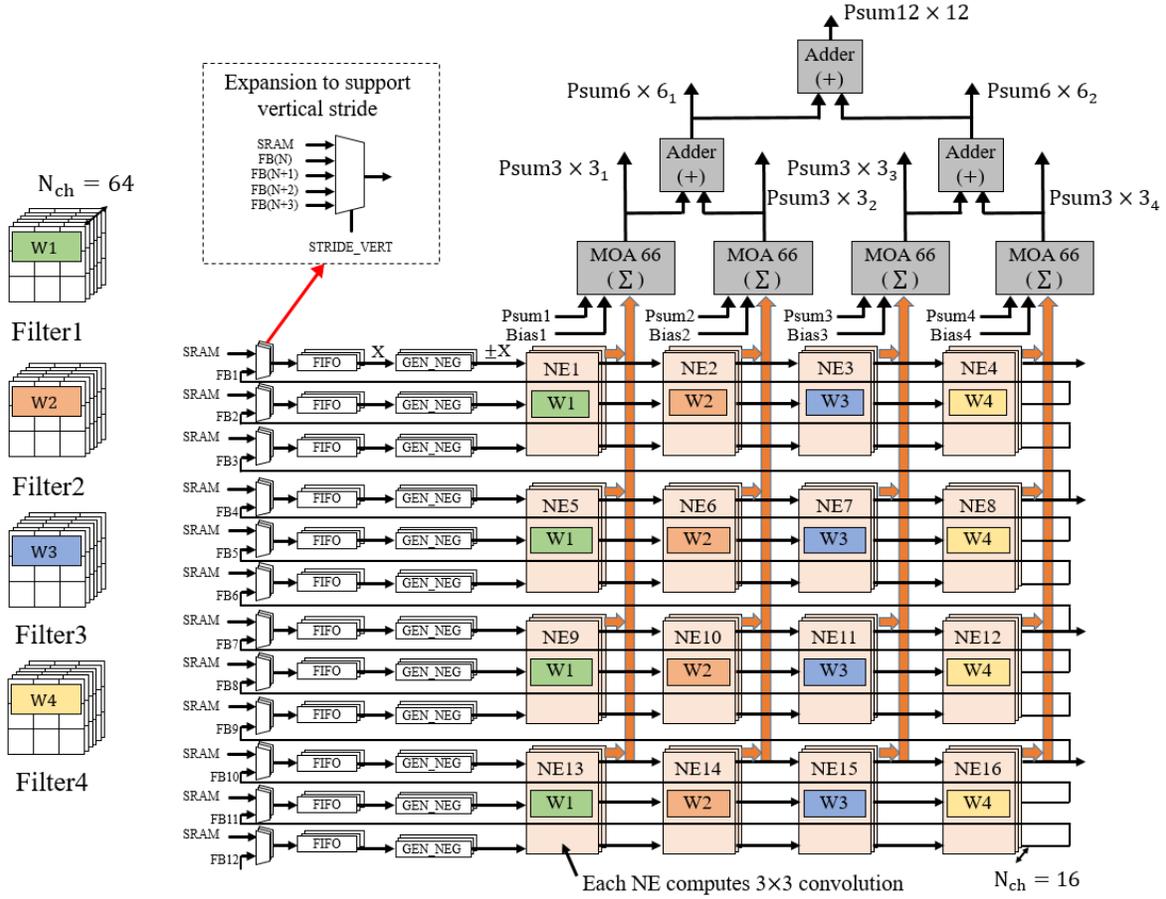

Fig. 5. Block diagram of the 4×4×16 NE array and its computation of four 3×3×64 convolutions.

NEs of one column. Therefore, in case of computing $3 \times 3 \times D$ convolution, four $3 \times 3 \times 64$ convolutions are computed in parallel. While sweeping $3 \times 3 \times H$ filters, inputs shifted out from the rightmost NEs are fed back into FIFO to reuse data, and therefore FIFO registers in only 3, 6, 9, and 12 rows receive input data from SRAM. GEN_NEG block generates a negatized input with a 2's complement circuit.

Three binary adders positioned in top of NE array are used in case that width or height of a filter is greater than 3, or in case that FC layer is computed. We will describe how to configure the filter size and compute FC layers in detail in the next subsection. If the mux receives FBs from the multiple rows as illustrated in box with dotted line in Fig. 5, the stride of convolution in vertical direction also can be configured.

Figure 6 shows block diagram of the system architecture of the proposed TMA accelerator. Inputs and weights stored in DRAM are transmitted to SRAM. The weights stored in SRAM are decomposed by the Weight decomposition block. The decomposed weights are loaded to Weight registers in 4×4×16 NE array. The Psums computed by the NE array are stored in address of the range called Psum of Layer N in SRAM, if they are Psums of Nth layer. If the Psum is a final sum of Nth layers, they are stored in address of the range called Layer N after applying activation and pooling operations. Data stored in Inputs or Layer N in SRAM are loaded to FIFO. The final output of neural networks is delivered to DRAM.

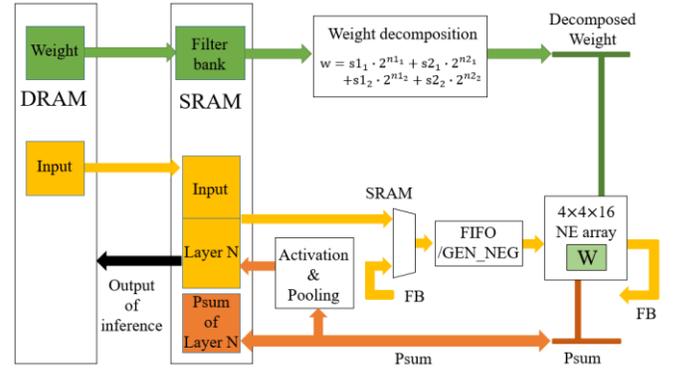

Fig. 6. Block diagram of the system architecture of the proposed TMA accelerator.

### C. filter size configuration and computation in FC layers

This subsection describes a configuration scheme of the proposed TMA accelerator that supports variable filter sizes and computation in FC layers.

Figure 7 illustrates a configuration scheme of the proposed TMA accelerator with example of convolution with $5 \times 5 \times D$ and $11 \times 11 \times D$, and computation in FC layers. In Case 1, assume that there are two $5 \times 5 \times D$ convolutional filters to compute. The proposed TMA assigns the weights of a $5 \times 5 \times 32$ subsection of the filter into two columns of the NE array. Zero bits are stored in the weight registers positioned at the right and bottom edges of the weight registers of the $2 \times 2$ NE array.

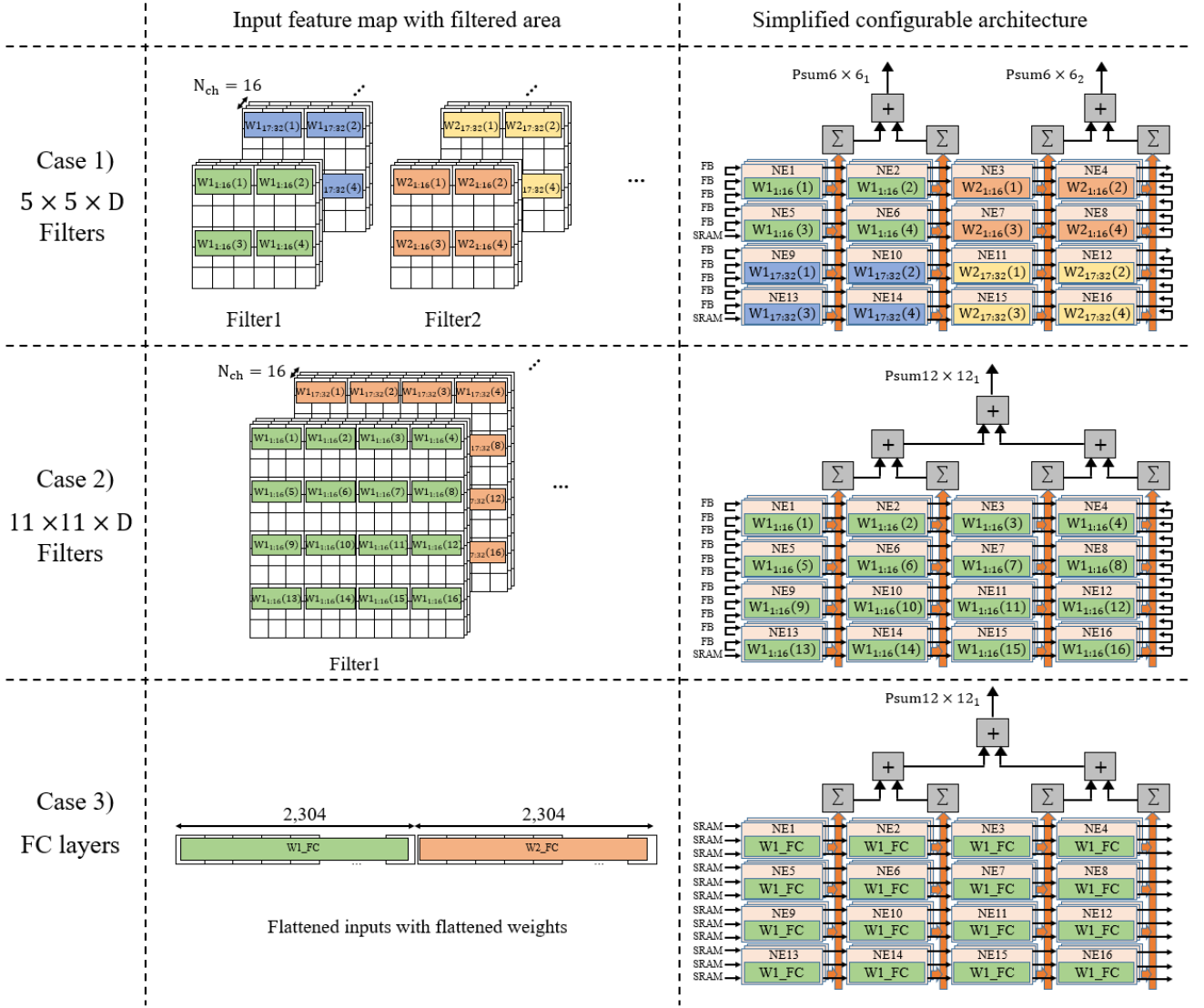

Fig. 7. Illustration describing configuration scheme of the proposed TMA accelerator with example of convolution with $5 \times 5 \times D$ and $11 \times 11 \times D$, and computation in FC layers.

Therefore, 6 rows with 16 channels of the input feature map are shifted through the two NE rows of the NE array. In this case, inputs in 10 rows are reused. The two Psum6x6s produce two 5×5×32 convolutions in parallel every input-shift.

In Case 2, suppose that we have a 11×11×D convolutional filter to compute. This time, weights of a 11×11×16 subsection of the filter are assigned into the entire 4×4×16 NE array. Zero bits are stored in the weight registers positioned at the right and bottom edges of the Weight registers of the 4×4 NE array. Then, 12 rows with 16 channels of the input feature map are shifted through the 4×4×16 NE array. In this case, inputs in 11 rows are reused. The Psum12×12 produces a 11×11×16 convolution every input-shift.

Lastly, in Case 3, 2,034 inputs are assigned in NE array per 12 input-shifts, and the top binary adder produces dot products of two vectors with 2,304 components is performed per 12 input-shifts. In this case, no rows are fed back.

## III. EXPERIMENTAL RESULTS

We implemented the proposed TMA in Xilinx Virtex-7 FPGA (XC7V2000T). Table II summarizes the performance of the implemented TMA accelerator. It employs 4×4×16 NE array as a computational engine (i.e., capacity of parallel 2,304 MACs). We set the capacity of the SRAM to 4 MB to reuse all Psums of all layers of AlexNet, where Block RAM (BRAM) provided in the FPGA is used as SRAM. The proposed architecture includes 12×16 FIFOs, where 12 and 16 are height and depth of NE array, respectively. We set the capacity of each FIFO to 224 Byte, since the widest width of input feature map among all Conv layers in AlexNet is 224. The number of the total gates used is ~294K. The peak throughput of the implemented TMA accelerator is achieved 576 GMACS in the case that bit-width of weight is 5-bit and 288 GMACS in the case of 8-bit. The frame rate for inference of AlexNet is 62 Frame/s at 200 MHz.

TABLE II
PERFORMANCE OF THE IMPLEMENTED TMA ACCELERATOR ON FPGA
XILINX XC7V2000T

| Performance of the implemented TMA accelerator | |
| --- | --- |
| Number of MACs | 2,304 |
| SRAM (BRAM in FPGA is used) | 4 MB |
| Clock frequency | 200 MHz |
| One FIFO capacity | 224 Byte |
| Peak throughput | 576 GMACS (INT5) 288 GMACS (INT8) |
| Gate count | 294 K |
| Inference frame rate (AlexNet) | 62 Frame/s |

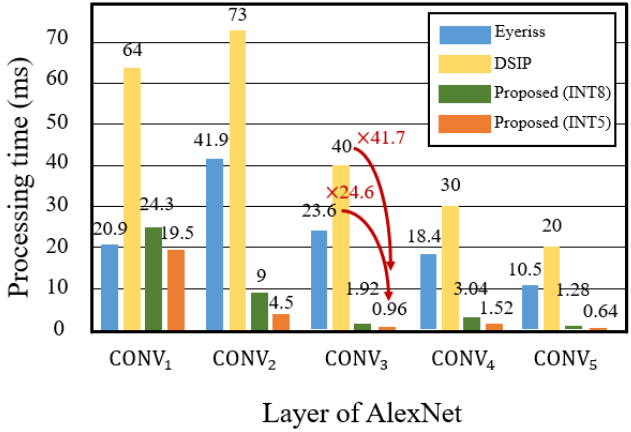

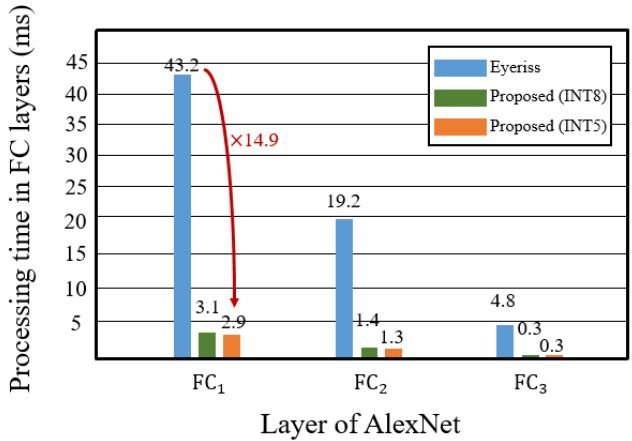

Fig. 8. Performance comparison of processing time of the proposed TMA accelerator with those in previous work in terms of computation of (a) Conv and (b) FC layers of AlexNet (batch = 4).

## IV. DISCUSSIONS

Discussions section describes structural advantages and improvements of the TMA accelerator over prior arts through three sub-sections. The three sub-sections are summarized as follows: improvement in processing time, SRAM access reduction, and improvement in MACS/W compared to prior

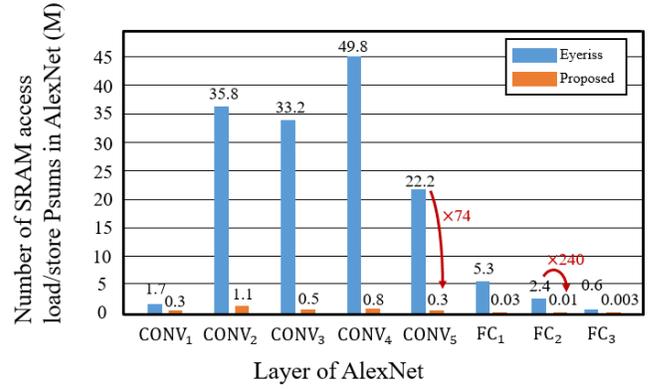

Fig. 9. Performance comparison of SRAM access of the proposed TMA accelerator with thous in previous work of AlexNet (batch = 1).

arts.

### A. Processing time of AlexNet

The reduction in circuit complexity of the proposed computational unit helps to enhance parallelism of computation, and allows to compute 2,304 MAC operations in parallel, while 168 and 64 MAC operations are computed in parallel in Eyeriss [5] and DSIP [8], respectively. The enhanced parallelism reduces the processing time of from Conv2 to Conv5 of AlexNet, compared to Eyeriss and DSIP, as shown in Fig 7 (a). Especially, in Conv3, the processing time of the proposed TMA with INT5 weights is reduced by ~24.6×  and ~41.7× , respectively, compared to Eyeriss and DSIP. However, in the case of the Conv1, the processing time of the proposed TMA with INT8 weights takes longer than that of Eyeriss. The reason is that only 11×11×3 SAMs among 12×12×16 SAMs is used, since the depth of Conv1 is 3.

The proposed TMA consumes additional cycles if accumulating PSIs. Therefore, processing with INT8 weights consumes approximately twice as many cycles as that with INT5 weights in from Conv2 to Conv5. However, in Conv1, the process with the INT8 weights consumes ~1.25× more cycles than that with INT5 weights. The reason is as follow: We did not implement the configuration of stride in horizontal direction in the proposed scheme. Therefore, in processing Conv1, the PSIs are accumulated after 4 Input-shifts, since the stride of Conv1 of AlexNet is 4.

Figure 8 shows improvement in processing time in FC layers of AlexNet compared to Eyeriss. The processing time in FC1 layer of the proposed TMA with INT8 and INT5 weights is reduced by ~13.9× and ~14.9×, respectively, compared to that of Eyeriss. In the case of FC layers, 2,304 MAC operations are performed per 12 Input-shifts. Therefore, clock cycles consumed in Input-shift is more dominant than that consumed in accumulation of PSIs. Thus, performance degradation caused by accumulation of PSIs is less than 10 % in the case of processing FC layers in AlexNet.

### B. Reduction for Storing and loading Psums in SRAM access

The proposed TMA delivers one, two, or four Psums to SRAM depending on the configuration, although it computes 2,304 MACs in parallel. On the other hand, Eyeriss transmits 12 Psums, while computing 168 MACs in parallel. Therefore, SRAM accesses required for loading and storing Psums is reduced although the proposed scheme increases parallelism of computation, resulting in reduction in power consumption and alleviation of the memory bottleneck. Fig. 9 shows the reduction in the number of SRAM accesses to load and store Psums in AlexNet. In Conv layers, it is reduced by ~74× in maximum and in FC layers, by ~240× in maximum.

TABLE III
PERFORMANCE COMPARISON OF THE PROPOSED TMA ACCELERATOR WITH THOSE IN PREVIOUS WORKS.
ASTERISK MARK INDICATES SIMULATED RESULT

|  | Eyeriss [5] | ConvNet [6] | DSIP [8] | This work |
| --- | --- | --- | --- | --- |
| Weight bit-width | 16-bit | 16-bit | 16-bit | 5-bit/8-bit |
| Activation bit-width | 16-bit | 16-bit | 16-bit | 8-bit |
| Number of MACs | 168 | 256 | 64 | 2,304 |
| Technology (nm) | 65 | 40 | 65 | 65 |
| Operating voltage (V) | 1.0 | 0.9 | 1.2 | 1.0 |
| Power consumption (mW) | 278 | 274 | 88.6 | 237 * |
| Frequency (MHz) | 250 | 204 | 250 | 250 * |
| Throughput (GMACs) | 23.1 | 52.2 | 30.1 | 576 (INT5) * <br> 288 (INT8) * |
| Throughput per Watt. (GMACs/W) | 83.1 | 190.6 | 136.8 | 2,430 (INT5) * <br> 1,215 (INT8) * |

*C. Performance comparison with previous works*

Table III shows the performance comparison of the propose TMA with those in previous works, Eyeriss [5], ConvNet [6], and DSIP [8]. The prior works in Table 2 adopts 16-bit weights and activations. Therefore, booth multipliers used in those works produce 9 partial products and aggregate them [20]. However, the proposed TMA employs scalable INT5/INT8 weights with accumulating two PSIs. The reduced circuit complexity of the proposed computational engine enhanced parallelism resulting in computing more MAC operations in parallel. The number of MAC operation computed in parallel of the proposed TMA is 9× higher than those of ConvNet.

The power consumption of the proposed TMA is simulated at 250 MHz, where simulation environment is 65 nm CMOS technology at operating voltage of 1.0 V. The simulated power while processing layer2 in AlexNet is achieved 237 mW. The peak throughput per Watt of the proposed TMA is achieved 2.43 TMACS/W and 1.215 TMACS/W in case that the bit-widths of the weight are INT5 and INT8, respectively. Those performances are improved by ~12.7× and ~6.4× compared to ConvNet, respectively.

V. CONCLUSION

This paper proposes a fully configurable Tera-MACS/W neural hardware inference accelerator with a multiplier-less massive parallel processor with 8-bit integer activations and scalable integer weights. The proposed TMA with INT5 weights achieves the throughput of 576 GMACS and 2.4 TMACS/W. It reduces processing time of from Conv2 to Conv5 layers of AlexNet with INT5 weight by ~12.4× and ~21.4 × compared to Eyeriss and DSIP accelerators, respectively. It also improves processing time of all FC layers of AlexNet with INT5 weight by ~14.9× compared to Eyeriss. In addition, it reduces SRAM access for storing and loading Psums by ~49.6× compared to Eyeriss.

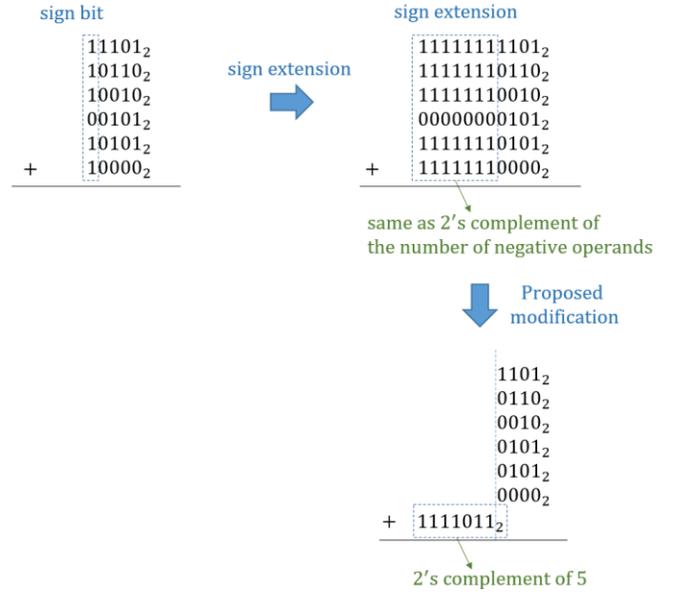

Fig. A1. An example of summation of six 5-bit binary numbers. This example describes summation of extended sign bits is same as 2's complement of the number of negative operands. Therefore, sign extension can be replaced to addition of one binary number.

APPENDIX

In Appendix, the principle of the proposed sign extension of the MOA circuit is described. Fig. A1 shows an example of summation of six 5-bit binary numbers. In the conventional way, negative numbers should extend 1s and positive numbers should extend 0s, as shown in Fig. A1. In a closer look, binary number of the extended 1s is same as -1 and that of the extended 0s is same as 0. Therefore, the summation of the extended bits can be replaced to 2's complement of the number of negative numbers.

ACKNOWLEDGMENT

"This research was supported by the MSIT(Ministry of Science and ICT), Korea, under the "ICT Consilience Creative Program" (IITP-2019-2017-0-01015) supervisedby the IITP(Institute for Information & communications

Technology Planning&Evaluation)"